\documentclass[twocolumn]{aastex62}

\shorttitle{H$\alpha$ in the Type~Ia SN2018cqj/ATLAS18qtd}
\shortauthors{Prieto et al.}

\begin{document}

\title{Variable H$\alpha$ Emission in the Nebular Spectra of the Low-Luminosity Type Ia SN2018cqj/ATLAS18qtd}

\author[0000-0003-1072-2712]{Jose~L.~Prieto}
\affil{N\'ucleo de Astronom\'ia de la Facultad de Ingenier\'ia y Ciencias, Universidad Diego Portales, Av. Ej\'ercito 441 Santiago, Chile}
\affil{Millennium Institute of Astrophysics, Santiago, Chile}

\author[0000-0003-0853-6427]{Ping Chen}\affil{Kavli Institute for Astronomy and Astrophysics, Peking University, Yi He Yuan Road 5, Hai Dian District, Beijing 100871, China.}\affil{Department of Astronomy, School of Physics, Peking University, Yi He Yuan Road 5, Hai Dian District, Beijing 100871, China}

\author[0000-0002-1027-0990]{Subo Dong},\affil{Kavli Institute for Astronomy and Astrophysics, Peking University, Yi He Yuan Road 5, Hai Dian District, Beijing 100871, China.}

\author[0000-0003-3529-3854]{S.~Bose}
\affil{Kavli Institute for Astronomy and Astrophysics, Peking University, Yi He Yuan Road 5, Hai Dian District, Beijing 100871, China.}

\author[0000-0002-3653-5598]{A.~Gal-Yam}
\affil{Benoziyo Center for Astrophysics and the Helen Kimmel center for planetary science, Weizmann Institute of Science, 76100 Rehovot, Israel}

\author[0000-0001-9206-3460]{T.~W.-S.~Holoien}
\affil{Observatories of the Carnegie Institution for Science, 813 Santa Barbara Street, Pasadena, CA 91101, USA}

\author[0000-0001-9852-1610]{J.~A.~Kollmeier}
\affil{Observatories of the Carnegie Institution for Science, 813 Santa Barbara Street, Pasadena, CA 91101, USA}

\author[0000-0003-2734-0796]{M.~M.~Phillips}
\affil{Las Campanas Observatory, Carnegie Observatories, Casilla 601, La Serena, Chile}

\author[0000-0003-4631-1149]{B.~J.~Shappee}
\affil{Institute for Astronomy, University of Hawai`i at Manoa, 2680 Woodlawn Dr., Honolulu, Hi 9682}

\begin{abstract}

We present optical photometry and spectroscopy of the Type~Ia supernova SN2018cqj/ATLAS18qtd. The supernova exploded in an isolated region at $\sim 65$~kpc from the S0 galaxy IC~550 at $z=0.0165$ ($D\approx 74$~Mpc) and has a redshift consistent with a physical association to this galaxy. Multicolor photometry show that SN2018cqj/ATLAS18qtd is a low-luminosity ($M_{B_{max}}\approx -17.9$~mag), fast-declining Type~Ia with color stretch $s_{BV} \approx 0.6$ and $B$-band decline rate $\Delta m_{15}(B) \approx 1.77$~mag. Two nebular-phase spectra obtained as part of the 100IAS survey at +193 and +307~days after peak show the clear detection of a narrow H$\alpha$ line in emission that is resolved in the first spectrum with $\rm FWHM \approx 1200$~km~s$^{-1}$ and $L_{H\alpha} \approx 3.8\times 10^{37}$~erg~s$^{-1}$.  
The detection of a resolved H$\alpha$ line with a declining luminosity is broadly consistent with recent models where hydrogen is stripped from the non-degenerate companion in a single-degenerate progenitor system. However, the amount of hydrogen consistent with the luminosities of the H$\alpha$ line would be $\sim 10^{-3}$~M$_{\odot}$, significantly less than theoretical model predictions in the classical single-degenerate progenitor systems. SN2018cqj/ATLAS18qtd is the second low-luminosity, fast-declining Type~Ia SN after SN2018fhw/ASASSN-18tb that shows narrow H$\alpha$ in emission in its nebular-phase spectra. 
\end{abstract}


\section{Introduction} 
\label{sec:intro}
Type~Ia supernovae (SNe) are among the most energetic and common stellar explosions, contribute to the nucleosynthesis of iron-peak elements, and have been used extensively to trace the expansion history of the Universe. However, the nature of their progenitors and explosion mechanisms is still unknown despite decades of detailed observations of hundreds of extragalactic Type~Ia SNe and several Galactic SN remnants. Multiple lines of observational evidence and theoretical modelling point to Type~Ia SNe as thermonuclear explosions of Carbon-Oxygen (CO) white dwarfs (WD), but the properties of the progenitor system (e.g., WD mass, degenerate or non-degenerate companion, single, binary or triple system) and the explosion mechanism (e.g., CO detonation/deflagration in an accreting WD, merger or collision of two WDs) are under debate \citep[e.g.,][and references therein]{wang12,hillebrandt13,maoz14,livio18}. 

In the single-degenerate (SD) progenitor channel, a CO WD accretes matter (hydrogen or helium rich) from a non-degenerate companion until it reaches a critical mass, typically the Chandrasekhar mass, and explodes \citep[e.g.,][]{han04,nomoto18}. One of the main predictions of the classical SD channel is that some amount of material ($M\sim 0.1-0.5$~M$_{\odot}$) is stripped from the non-degenerate companion by the SN shock and can be detected as narrow ($v\sim 1000-2000$~km~s$^{-1}$) emission lines from hydrogen or helium in late-time ($\gtrsim 200$~days after peak), nebular-phase spectra when the ejecta are optically thin \citep[e.g.,][]{wheeler75,chugai86,marietta00,liu12,boehner17,botyanszki18}. Modifications to the classical SD channel can lead to the companion losing significant mass before the SN explosion, potentially hiding these signatures \citep{justham11,distefano11}.

Several observational campaigns have targeted nearby Type~Ia SNe looking for signatures of the stripped material, mainly searching for an hydrogen H$\alpha$ line in deep nebular-phase optical spectra \citep[e.g.,][]{mattila05,leonard07,lundqvist13,shappee13,lundqvist15,maguire16,graham17,shappee18,sand18,holmbo19,dimitriadis19,tucker19a,sand19,tucker19b}, but have only obtained non-detections.

Recently, \citet{kollmeier19} reported the discovery of a strong H$\alpha$ line in emission with an integrated luminosity of $L\approx 2.2\times 10^{38}$~erg~s$^{-1}$ and line width of $\rm FWHM \approx 1100$~km~s$^{-1}$ in the +139~days nebular spectrum of the fast declining (color stretch $s_{BV}\approx 0.5$ and $\Delta m_{15}(B)\approx 2.0$~mag), low-luminosity ($M_{B_{max}} \approx -17.6$~mag) Type~Ia SN2018fhw/ASASSN-18tb \citep{brimacombe18}. If this H$\alpha$ detection is stripped hydrogen from the non-degenerate companion in an SD progenitor system, the amount of material needed to explain the line luminosity would be $\sim 2\times10^{-3}$~M$_{\odot}$. \citet{vallely19} presented more spectra, photometry, and an early rising light curve of SN2018fhw/ASASSN-18tb that put into question this interpretation of the observations and instead favor a scenario similar to Type~Ia-CSM events \citep[e.g.,][see also \citealt{graham19}]{hamuy03,prieto07,dilday12,silverman13}, where a strong H$\alpha$ emission line and extra luminosity most likely originate in the interaction between the SN ejecta and a hydrogen rich circumstellar medium (CSM). 

\begin{figure*}[ht!]
\begin{center}
\plotone{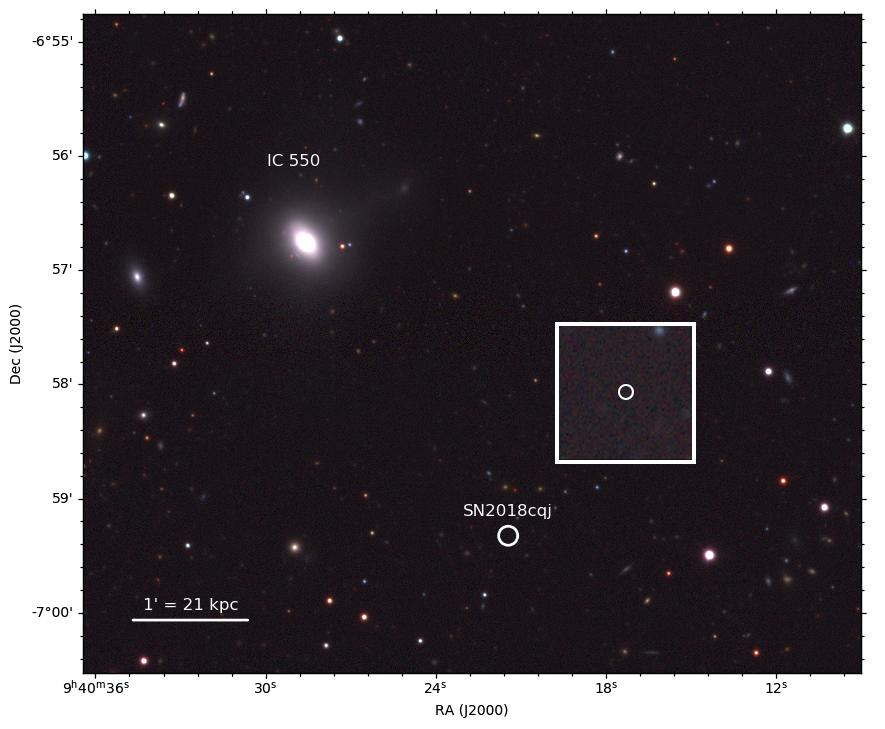}
\caption{DECals DR7 $grz$ deep color composite of the field of the Type~Ia SN2018cqj/ATLAS18qtd and the nearby S0 galaxy IC~550 ($z=0.016455$, $D\simeq 74$~Mpc). SN2018cqj/ATLAS18qtd is located at a projected separation of 3.2$\arcmin$ (65~kpc) from IC~550. The box shows a zoom-in to a $20\arcsec\times 20\arcsec$ region around the position of SN2018cqj/ATLAS18qtd, marked with a $1\arcsec$ radius circle.\label{fig1}}
\end{center}
\end{figure*}

In this paper, we present observations of the Type~Ia SN2018cqj/ATLAS18qtd that show similarities to SN2018fhw/ASASSN-18tb, including the detection of an H$\alpha$ emission line in late-time spectra of a low-luminosity SN~Ia. The spectroscopic observations of SN2018cqj/ATLAS18qtd presented in this paper were obtained as part of the 100IAS survey \citep{dong18,kollmeier19,chen19}, a project using 5-10 meter class telescopes to obtain nebular-phase optical spectra of a complete sample of $\sim 100$ SNe~Ia within $z\lesssim 0.02$, in order to systematically study signatures of explosion asymmetries such as those discovered in SN2007on \citep{dong15}. In Section~\S\ref{sec:env} we discuss the discovery, spectroscopic classification, and host galaxy environment of SN2018cqj/ATLAS18qtd. In Section~\S\ref{sec:phot} and \S\ref{sec:spec} we present the early and late-time optical light curve and spectra, respectively. In Section~\S\ref{sec:discussion} we present a discussion of the results and conclusions.

\section{Data and Analysis} 
\label{sec:data}

\subsection{Discovery, Classification, and Host Environment}
\label{sec:env}
SN2018cqj/ATLAS18qtd was discovered by the Asteroid Terrestrial-impact Last Alert System \citep[ATLAS;][]{atlas} transient survey at RA = 09:40:21.463 and DEC = $-$06:59:19.76 (J2000.0) on 2018 June 13.27 (all dates are UT) with orange magnitude $o = 18.14 \pm 0.09$~mag and had a pre-discovery non-detection on 2018 June 11.25 with $o > 18.58$~mag \citep{tonry18_disc}. After initial inspection of archival imaging data from the Digital Sky Survey and Pan-STARRS1 \citep[PS1;][]{ps1_1,ps1_2}, we found no obvious host galaxy at the location of the transient. However, the NASA Extragalactic Dabase (NED) shows that the transient was located 3.1$\arcmin$ from the nearby S0 galaxy IC~550 at $z=0.016455 \pm 0.000150$ \citep{hyperleda,6dF}, potentially making it a low-redshift supernova and worth following up (see Figure~\ref{fig1}). 

An optical low-resolution spectrum obtained by \citet{khlat18} on 2018 July 14.97, presented in Section~\S\ref{sec:spec}, showed features characteristic of a Type~Ia supernova a few weeks after maximum at $z\approx 0.02$, making a physical association with the S0 galaxy IC~550 possible. We re-analyzed the spectrum using the Supernova Identification cross-correlation code \citep[SNID;][]{snid}. Fitting the spectrum with SNID in the wavelength range $4200-7200$~\AA, cutting the low signal-to-noise ratio ends of the spectrum, we find that it is most consistent with a normal Type~Ia SN at $31 \pm 7$~days after maximum light and with a redshift $z_{SN}=0.0155 \pm 0.0019$. 

\begin{figure*}
\epsscale{1.2}
\plotone{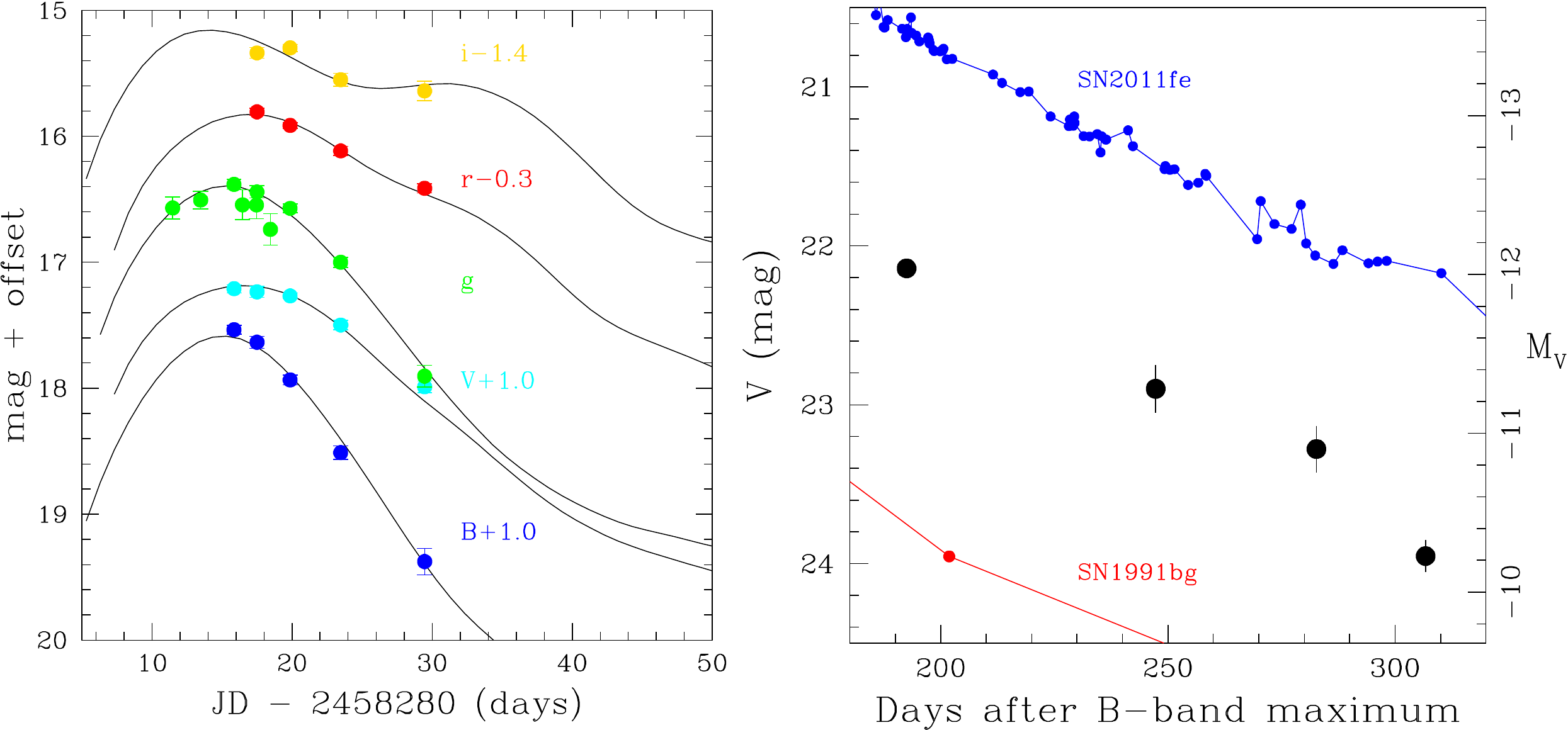}
\caption{{\it Left:} $BVgri$ light curves around peak of the Type~Ia SN2018cqj/ATLAS18qtd. The lines are the best fit SNooPy \citep{snoopy} Type Ia SN light curve template. We have applied an arbitrary offset in magnitude to each band, shown in the labels. {\it Right:} Late-time $V$-band light curve of SN2018cqj/ATLAS18qtd (filled black circles). For comparison, we also show the late-time light curves of the normal Type Ia SN2011fe \citep[blue points and line;][]{munari13} and the low peak luminosity, fast declining Type~Ia SN1991bg \citep[red point and line;][]{turatto96}. \label{fig2}}
\end{figure*}

The redshift derived with SNID from the supernova features is fully consistent, within the 1$\sigma$ uncertainties, with the redshift of IC~550, which makes a physical association between SN2018cqj/ATLAS18qtd and the S0 galaxy likely. Throughout the rest of the paper, we assume that IC~550 is the host galaxy of SN2018cqj/ATLAS18qtd, which at  $z_{CMB}=0.017596$ (from the NED) gives a luminosity distance of $D_L=74.3$~Mpc with an assumed Hubble constant of $H_0=72$~km~s~Mpc$^{-1}$. At this distance, the projected physical separation between the supernova and the center of IC~550 is $\sim 65$~kpc, which would imply that it went off in the outer halo of the S0 galaxy. However, we cannot rule out that the host is a smaller dwarf galaxy closer to the supernova in a galaxy group with IC~550. 
Figure~1 shows a color composite image of the environment of SN2018cqj/ATLAS18qtd obtained using the deepest archival images of the field from the Dark Energy Camera Legacy Survey \citep[DECals;][]{decals}. The DECals DR7 $grz$ coadds give strong non-detection upper limits at the position of the supernova of $g>25.8$~mag ($M_g > -8.7$), $r> 24.6$~mag ($M_r > -9.8$) and $z>23.4$~mag ($M_z > -11.0$) at 5$\sigma$, which rule out an association of the progenitor system with dwarf galaxies down to the luminosities of massive globular clusters \citep{mcconnachie12}. 

\subsection{Photometry}
\label{sec:phot}
We obtained five epochs of photometric follow-up observations of SN2018cqj/ATLAS18qtd, close to its peak magnitude, in the $BVgri$ bands between 2018 June 26.4 and July 10.0 with the Las Cumbres Observatory Global Telescope Network \citep[LCOGT;][]{lcogt} 1m telescopes at the Cerro Tololo Interamerican Observatory (CTIO) and the Siding Spring Observatory. We ran Point-Spread-Function (PSF) photometry on the LCOGT pipeline-reduced images using the DoPHOT package \citep{dophot1,dophot2}. We calibrated the $BVgri$ magnitudes using the AAVSO Photometric All-Sky Survey \citep[APASS;][]{apass}. 

We also retrieved five epochs of $g$-band photometry obtained between 2018 June 22.0 and June 29.0 by the All-Sky Automated Survey for SuperNovae \citep[ASAS-SN][]{asassn,kochanek17} from the Bohdan Paczy\'nski unit in CTIO. The ASAS-SN images were reduced with an automated pipeline using the ISIS image subtraction package \citep{isis1,isis2} and the photometry was calibrated with APASS stars in the field.  

Additionally, we obtained four late-time ($>190$ days after peak) $V$-band photometric measurements of SN2018cqj/ATLAS18qtd between 2019 January 7.4 and 2019 May 3.5 from images obtained with the FORS2 instrument \citep{fors2} on the VLT UT1 Antu 8.2m telescope at the ESO Paranal Observatory and with the WFCCD instrument on the du~Pont 2.5m telescope at the Las Campanas Observatory. The images were reduced using standard techniques in IRAF. PSF photometric measurements of the SN were extracted using DoPHOT and calibrated with APASS and PS1 stars in the field. 

We present the photometry of SN2018cqj/ATLAS18qtd in Table~\ref{table:tab1}. Figure~\ref{fig2} shows the early (left panel) and late-time (right panel) light curve evolution of SN2018cqj/ATLAS18qtd. We fitted the $BVgri$ light curves around peak using the SNooPy fitting package \citep{snoopy}. We choose the maximum light models with the color stretch parameter $s_{BV}$, which is less sensitive to having poorly sampled light curves before peak brightness. The results obtained with the SNooPy light curve fits are: $t_{max} = 2458295.2\pm 0.9$~days (2018 June 25.7), $s_{BV} = 0.61 \pm 0.06$, $B_{max} = 16.46 \pm 0.08$~mag, $V_{max} = 16.08\pm 0.06$~mag, $g_{max}=16.29\pm 0.02$~mag, $r_{max}=16.10\pm 0.05$~mag, and $i_{max}=16.52\pm 0.11$~mag, where the magnitudes at maximum have been corrected for Galactic extinction and $K$-corrections. 

\begin{figure*}
\begin{center}
\epsscale{1.2}
\plotone{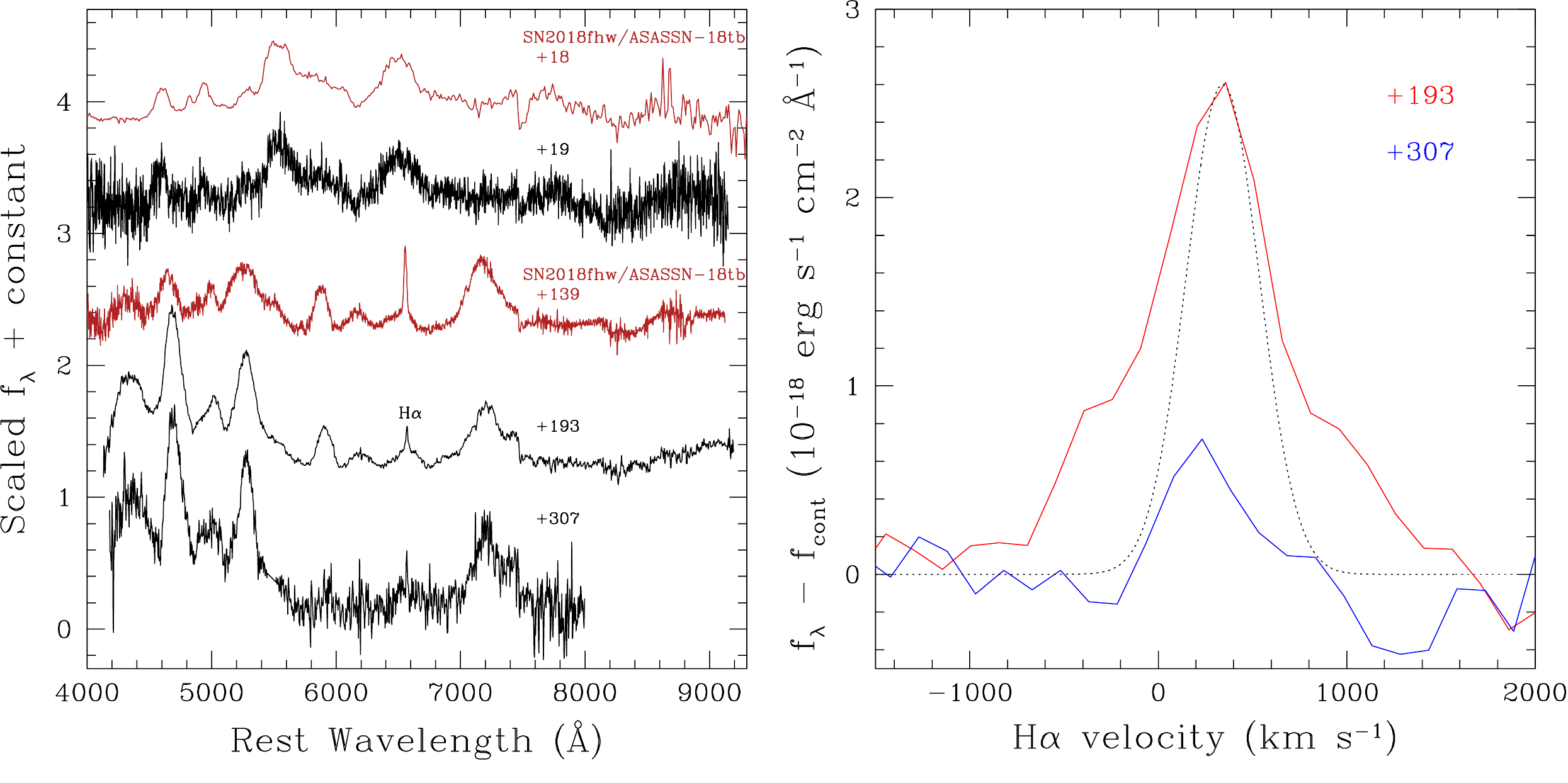}
\caption{{\it Left:} Optical spectral sequence of the Type~Ia SN2018cqj/ATLAS18qtd in black. We include the +18~days and the +139~days nebular-phase spectrum of the Type~Ia SN2018fhw/ASASSN-18tb in red \citep{vallely19,kollmeier19}. All the spectra have been scaled and offset in flux for clarity. We show the location of the H$\alpha$ emission line in the +193~days nebular-phase spectrum of SN2018cqj/ATLAS18qtd. The phases of the spectra, in days after peak in $B$-band, are also shown. {\it Right:} H$\alpha$ emission line profiles in the two nebular-phase spectra of SN2018cqj/ATLAS18qtd at +193 and +307~days after peak. We have subtracted the underlying continuum and corrected for Galactic extinction. The dotted line shows a Gaussian profile with the FWHM of the skylines (instrumental profile), shifted in velocity and scaled to the peak of the H$\alpha$ line at +193~days. \label{fig3}}
\end{center}
\end{figure*}

We derive a $B$-band magnitude decline rate at 15 days after peak of $\Delta m_{15}(B) = 1.77 \pm 0.10$~mag using the relation between $s_{BV}$ and  $\Delta m_{15}(B)$ in \citet{burns18}. The color stretch and decline rate, the $B$-band absolute magnitude at peak of $M_{B_{max}}=-17.9 \pm 0.2$~mag, and the red color at peak of $(B_{max}-V_{max}) = 0.4 \pm 0.1$~mag, are consistent with low-luminosity, fast-declining Type~Ia supernovae \citep{gall18,burns18}. 

The late-time $V$-band light curve of SN2018cqj/ATLAS18qtd is presented in the right panel of Figure~\ref{fig2} compared with the light curves of the normal Type~Ia SN2011fe \citep[$\Delta m_{15}(B)=1.11$~mag and $M_{B_{max}}= -19.2$~mag;][]{munari13} and the low-luminosity, very rapidly declining Type~Ia SN1991bg \citep[$\Delta m_{15}(B)=1.93$~mag and $M_{B_{max}}= -16.7$~mag;][]{turatto96,phillips99}. For direct comparison, we place the magnitudes of SN2011fe and SN1991bg at the distance and extinction of SN2018cqj/ATLAS18qtd using the derived distances to their host galaxies \citep{shappee11,cantiello11}. The $V$-band decline rate of SN2018cqj/ATLAS18qtd at $\sim 200-300$~days after peak is 0.015~mag~day$^{-1}$, fully consistent with the decline rates of SN2011fe and SN1991bg at the same late-time epochs. 

\subsection{Spectroscopy}
\label{sec:spec}

We obtained three optical long-slit spectra of SN2018cqj/ ATLAS18qtd\footnote{The spectra will be made publicly available through the Weizmann Interactive Supernova data REPository \citep[WISeREP;][]{yaron12}.}. Table~\ref{table:tab2} has a summary of the observations. The classification spectrum was obtained on 2018 July 14.97 with the WFCCD instrument on the du~Pont 2.5m telescope at Las Campanas Observatory. Two sets of nebular-phase spectra were obtained with the FORS2 instrument on the VLT UT1 telescope. The first set of five exposures were obtained as part of the period 102 regular ESO program 0102.D-0287(A) between 2019 January $5.2-9.2$, with a total exposure time on source of 7040~sec. The second set of four exposures were obtained as part of the period 103 DDT program number 2103.D-5008(A) between 2019 May $3.1-5.0$, with a total exposure time on source of 5800~sec. 

Standard reductions of all the 2D spectral images up to flat-fielding were obtained with tasks in IRAF {\it ccdproc}. The wavelength calibration, 1D spectral extraction, and flux calibration of each individual spectrum were done with tasks in {\it longslit}, {\it twodspec}, and {\it onedspec} packages in IRAF. For the nebular-phase spectra, we combined the fully reduced 1D spectra in each of the two sets of exposures obtained in January and May 2019, respectively. We checked and applied small offsets ($\lesssim 1$~\AA) to the wavelength calibration of each individual spectrum using the strong [\ion{O}{1}] sky emission line at $5577.3$~\AA. In order to obtain good absolute flux calibration of the final nebular-phase spectra to measure emission line fluxes, we obtained synthetic $V$-band magnitudes from the spectra and applied multiplicative factors to match the $V$-band photometry measured from the acquisition images (listed in Table~\ref{table:tab1}). 

The spectra of SN2018cqj/ATLAS18qtd are shown in Figure~\ref{fig3}. In the left panel of the figure we show the classification spectrum and the two nebular-phase spectra scaled and offset in flux (arbitrarily) for clarity. The wavelengths in the spectra are corrected to the rest-frame using the redshift of IC~550. The two nebular-phase spectra show the clear detection of a narrow, but resolved, emission line approximately at the wavelength of the H$\alpha$ transition. In the same panel, we show the +18~days and the +139~days nebular-phase spectrum of SN2018fhw/ASASSN-18tb for comparison \citep{vallely19,kollmeier19}. The nebular-phase spectrum of SN2018fhw/ASASSN-18tb has a significantly stronger intrinsic H$\alpha$ emission line with respect to the local continuum and also in peak/integrated flux (as we show below). 

In the right panel of Figure~\ref{fig3} we show the profiles of the H$\alpha$ line in the two nebular-phase spectra of SN2018cqj/ATLAS18qtd. We corrected the fluxes in each spectrum for Galactic extinction \citep{schlafly11}, and fitted and subtracted a 3rd order polynomial to the continuum around the narrow H$\alpha$ emission line. The x-axis in the panel shows the velocity with respect to the H$\alpha$ wavelength in the laboratory (6562.82~\AA) after correcting the wavelengths to the rest-frame using the redshift of IC~550. We also show a Gaussian profile with the FWHM of the skylines (see Table~\ref{table:tab2}) scaled to the central velocity and peak of the H$\alpha$ emission line in the +193~days spectrum. This shows that the H$\alpha$ emission line is clearly resolved with respect to the instrumental profile in the +193~days spectrum.  

The H$\alpha$ profile of the +193~days spectrum is best fit with a two-component Gaussian profile with the following best-fit parameters: $F_{1} = (4.55 \pm 0.40) \times 10^{-17}$~erg~s$^{-1}$~cm$^{-2}$, $\lambda_{1}= 6568.72 \pm 0.73$~\AA, $\rm FWHM_{1}=  28 \pm 3$~\AA\ and $F_{2} = (1.13 \pm 0.38) \times 10^{-17}$~erg~s$^{-1}$~cm$^{-2}$, $\lambda_{2} = 6569.92 \pm 0.49$~\AA, $\rm FWHM_{2}=  9 \pm 2$~\AA, where $F$ is the integrated line flux, $\lambda$ is the rest-frame wavelength at peak, and FWHM is the full-width at half maximum of the Gaussian profile in the rest-frame. The H$\alpha$ profile of the +307~days spectrum is best fit with a single Gaussian with best-fit parameters: $F = (6.96 \pm 1.55) \times 10^{-18}$~erg~s$^{-1}$~cm$^{2}$, $\lambda = 6567.94 \pm 0.99$~\AA, $\rm FWHM = 9 \pm 2$~\AA. 

The total integrated fluxes and luminosities of the H$\alpha$ line in both nebular-phase epochs are: $F(193) = (5.7\pm 0.6)\times 10^{-17}$~erg~s$^{-1}$~cm$^{-2}$, $L(193)=(3.8\pm 0.9)\times 10^{37}$~erg~s$^{-1}$ and $F(307) = (7.0\pm 1.7)\times 10^{-18}$~erg~s$^{-1}$~cm$^{2}$, $L(307)=(4.6\pm 1.4)\times 10^{36}$~erg~s$^{-1}$. In the flux uncertainties we have included an overall flux-calibration uncertainty given by the errors in the magnitudes in Table~\ref{table:tab1} and we have also added a 10\% uncertainty in the luminosity distance.
  
\section{Discussion and Conclusions} 
\label{sec:discussion}

We have presented optical photometry, spectroscopy, and archival imaging of SN2018cqj/ATLAS18qtd that show this transient is a fast declining (color stretch $s_{BV}\approx 0.6$ and $B$-band decline rate $\Delta m_{15}(B) \approx 1.77$~mag), low peak luminosity ($M_{B_{max}}\approx -17.9$~mag) Type~Ia supernova that exploded in the outer halo of the S0 galaxy IC~550 ($D\approx 74$~Mpc) at a projected separation of $\sim 65$~kpc or in an unidentified dwarf galaxy in the same galaxy group. Nebular-phase spectra obtained with the VLT~FORS2 instrument at +193 and +307~days after peak brightness show the detection of a strong, narrow line in emission associated with the supernova. The emission line is consistent with H$\alpha$ at a velocity of $\sim 300$~km~s$^{-1}$ with respect to the recession velocity of IC~550.  

SN2018cqj/ATLAS18qtd is the second case of a fast declining, low-peak luminosity Type~Ia SN observed by the 100IAS survey with an intrinsic H$\alpha$ line in emission detected in its nebular-phase spectra after SN2018fhw/ASASSN-18tb \citep{kollmeier19,vallely19}. \citet{kollmeier19} discussed the possibility that the H$\alpha$ detected in SN2018fhw/ASASSN-18tb could be stripped hydrogen from a hydrogen-rich non-degenerate companion in a single-degenerate progenitor system and estimated a stripped mass of $\sim 2\times 10^{-3}$~M$_{\odot}$ using the models of \citet{botyanszki18}. The amount of hydrogen could be as high as $\sim 10^{-2}$~M$_{\odot}$ taking into account the lower Ni yield of SN2018fhw/ASASSN-18tb ($\sim 0.1-0.2$~M$_{\odot}$) compared to normal Type~Ia SNe. Other possibilities for the origin of the H$\alpha$ line discussed by \citet{kollmeier19} included interaction between the SN ejecta and the CSM of the progenitor system, akin to the luminous Type~Ia-CSM events \citep[e.g.,][]{hamuy03,prieto07,dilday12,silverman13}, and fluorescent UV pumping in a slowly expanding shell of material. 

The detailed properties of the line emission in SN2018fhw/ASASSN-18tb $-$ the H$\alpha$ luminosity did not show significant variations (within a factor of 2) in spectra obtained between +37 and +148~days after peak and the non-detection of the H$\beta$ line which implied a high Balmer decrement of $F(H\alpha)/F(H\beta) \gtrsim 6$, and the smooth early light curve rise observed by TESS $-$ led \citet{vallely19} to suggest that the H$\alpha$ emission line originated from ejecta-CSM interaction instead of from stripped hydrogen in a single-degenerate system.

\begin{figure*}
\begin{center}
\epsscale{0.9}
\plotone{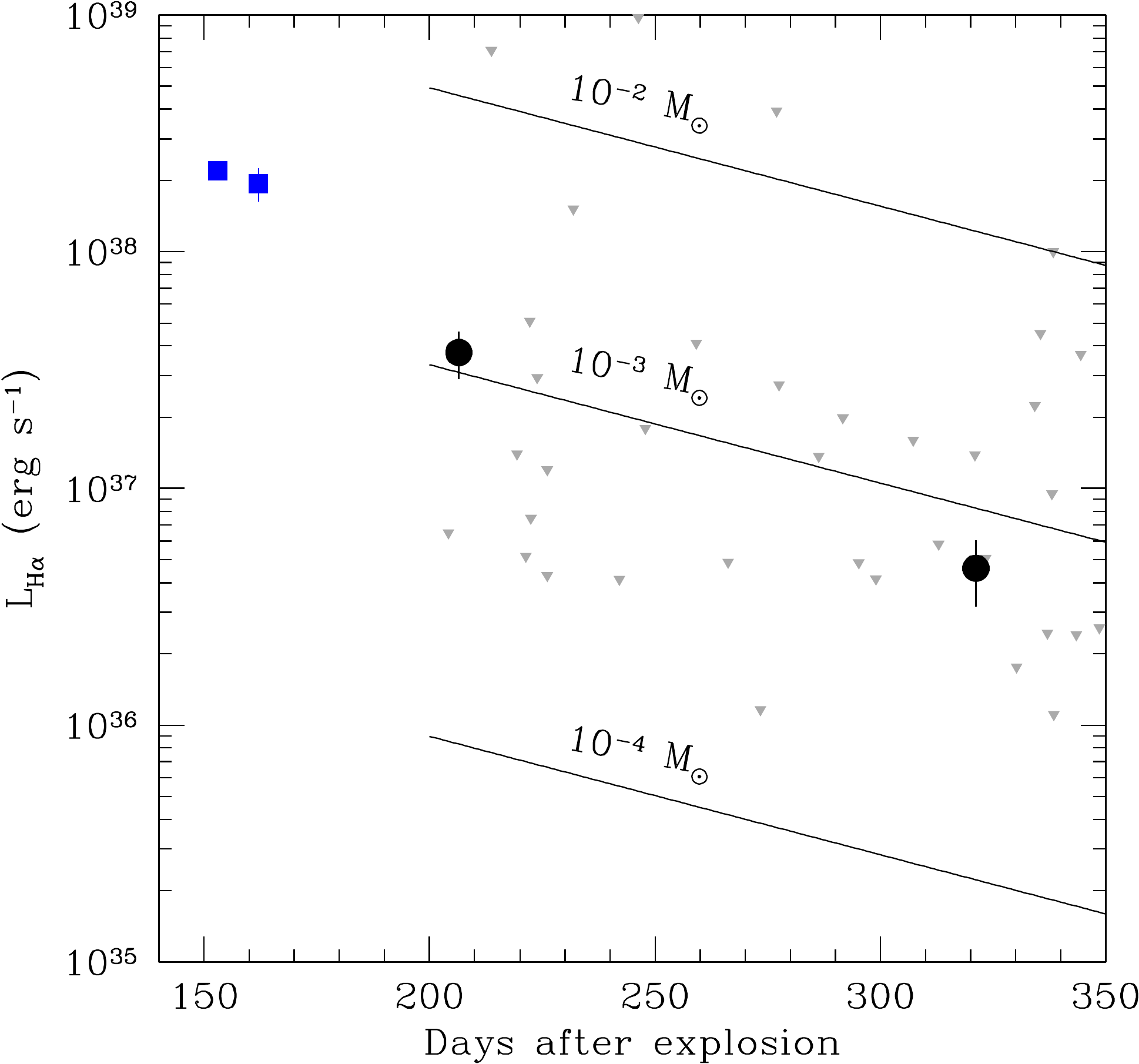}
\caption{Evolution of the H$\alpha$ emission line luminosity in the nebular-phase spectra of the Type~Ia SN2018cqj/ATLAS18qtd (filled dots). The lines show the evolution of the H$\alpha$ line luminosity predicted from models where hydrogen is stripped from the companion in a single degenerate progenitor \citep{botyanszki18,tucker19b}, which are valid at $>200$~days. The total amount of stripped hydrogen mass is shown above each line. We also show the luminosity of the H$\alpha$ line detected in the nebular-phase spectra of SN2018fhw/ASASSN-18tb  \citep[filled squares;][]{kollmeier19,vallely19} and 10$\sigma$ upper limits for 58 Type~Ia SNe  \citep[filled triangles;][]{tucker19b}.  \label{fig4}}
\end{center}
\end{figure*}

As in SN2018fhw/ASASSN-18tb, the detailed properties of the late-time emission observed in SN2018cqj/ ATLAS18qtd can shed light on the origin of the H$\alpha$ line detected in the two nebular-phase spectra. The H$\alpha$ emission line is clearly resolved in the +193~days spectrum, with a FWHM of the broad Gaussian component of $\approx 1200$~km~s$^{-1}$ after correcting for the intrinsic spectral resolution listed in Table~\ref{table:tab2} ($\sim 450$~km~s$^{-1}$), but is unresolved in the +307~days spectrum. However, we note that the lower signal-to-noise ratio of the second nebular spectrum would in part hinder the detection of a broad component. The total integrated luminosity of the H$\alpha$ line decreases by a factor of $\sim 8$ between the +193 and +307~days spectra. On the other hand, the $V$-band continuum magnitude is intermediate between SN2011fe (normal SN~Ia) and SN1991bg (fast decliner) with a consistent late-time decline rate of 0.015~mag~day$^{-1}$, as shown in Figure~\ref{fig2}, and the $V$-band flux decreases by a factor of $\sim 5$ between these epochs. We also searched for the H$\beta$ line in the +193~days spectrum and do not detect it. Using the method outlined in \citet{tucker19b}, we obtain a 10$\sigma$ upper limit on the H$\beta$ flux of $F(H\beta) \approx 10^{-17}$~erg~s$^{-1}$~cm$^{-2}$ (extinction corrected), which implies a lower limit on the Balmer decrement of $F(H\alpha)/F(H\beta) \gtrsim 6$, similar to SN2018fhw/ASASSN-18tb. 

In summary, most of the properties of the late-time emission in the fast declining, low-luminosity Type~Ia SN2018cqj/ATLAS18qtd are broadly consistent with the signatures expected when hydrogen is stripped from the non-degenerate companion star in a single-degenerate progenitor system \citep[e.g.,][]{botyanszki18,tucker19b}: the detection of a resolved H$\alpha$ line with $v \sim 1000$~km~s$^{-1}$ with an integrated luminosity that decreases in time roughly following the luminosity evolution of the supernova. Also, in the case of SN2018cqj/ATLAS18qtd the $V$-band light curve behaves like other normal and fast declining Type~Ia SNe at $\sim 200-300$~days after peak, and it does not show significant extra luminosity from ejecta-CSM interaction as it has been clearly observed in Type~Ia-CSM events \citep{silverman13}. The main property that is shared with Type~Ia-CSM events and also SN2018fhw/ASASSN-18tb is the high value of the Balmer decrement, which points to either high host galaxy extinction or that the origin of the emission line is different from the classical Case-B recombination with a Balmer decrement of $2.86$ \citep{osterbrock06}. We can, however, rule out significant host galaxy extinction because of the host galaxy environment of the SN and also because the shape of the continuum of the nebular-phase spectra is consistent with low-extinction Type~Ia SNe. 

We can use the luminosities of the H$\alpha$ lines in the nebular spectra of SN2018cqj/ATLAS18qtd to constrain the amount of hydrogen stripped off the putative companion to the white dwarf in the single-degenerate scenario interpretation. In Figure~\ref{fig4} we show the evolution of the H$\alpha$ luminosity as a function of time since explosion (filled circles). We have assumed a $B$-band rise-time from explosion to maximum light of $14$~days for fast declining Type~Ia SNe \citep{zheng17}. The three lines in Figure~\ref{fig4} have a constant amount of stripped hydrogen of $M_H= 10^{-4}, 10^{-3}, 10^{-2}$~M$_\odot$ and were obtained from the models of \citet{botyanszki18}, adapted in Equation~1 of \citet{tucker19b} to include the luminosity evolution of Type~Ia SNe between $200-500$~days after explosion. From the H$\alpha$ luminosities measured in the two nebular-phase spectra of SN2018cqj/ATLAS18qtd, we get a stripped hydrogen mass of $M_H = (1.2\pm 0.2) \times 10^{-3}$~M$_\odot$ and $(0.7\pm 0.2)\times 10^{-3}$~M$_\odot$. 

We can also use the linear relation between the peak luminosity of the nebular H$\alpha$ line and the amount of stripped material derived by \citet{mattila05}, as shown in \citet{leonard07}. From their relation, we get stripped hydrogen masses of $M_H\sim 0.27$~M$_\odot$ and $\sim 0.07$~M$_\odot$ from the two nebular-phase spectra of SN2018cqj/ATLAS18qtd. The \citet{mattila05} models have some limitations compared to the \citet{botyanszki18} models: they use spherically symmetric radiate transfer calculations, assume a uniform density of Solar-abundace material at the center of the ejecta up to a fixed velocity of $1000$~km~s$^{-1}$, and do not carry out time-dependent calculations. These are the main reasons we use to highlight the estimates obtained from the \citet{botyanszki18} models. However, despite the more realistic models, we note the potential for systematic uncertainties introduced due to the theoretical limitations of \citet{botyanszki18}. More realistic estimates will need a broader range of binary models, hydrodynamics models, and fuller treatment of atomic processes and radiative transfer. 

The H$\alpha$ emission line detected in the two nebular-phase spectra of SN2018cqj/ATLAS18qtd shows a redshift of $\sim 300$~km~s$^{-1}$ with respect to the recession velocity of the S0 galaxy IC~550. In the single-degenerate scenario, there is a viewing angle dependence in the material stripped off the companion and shifts in the centroid of the H$\alpha$ line of $\sim 10$~\AA\ ($\pm 500$~km~s$^{-1}$) from the rest-wavelength can be expected \citep{botyanszki18}. This velocity shift could also be caused by a different intrinsic velocity of the progenitor of SN2018cqj/ATLAS18qtd with respect to the central regions of IC~550, either as a fast-moving star in the outer halo of the galaxy (the central velocity dispersion of IC~550 is $\sigma = 158 \pm 15$~km~s$^{-1}$, \citealt{campbell14}) or if it comes from a different host galaxy in the same galaxy group as IC~550. Another possibility is that the emission line we detect is not H$\alpha$, but a transition from a different ion. \citet{jerkstrand15} and \citet{fangmaeda18} have interpreted the H$\alpha$-like emission detected in the nebular-phase spectra of Type~IIb SNe as the [\ion{N}{2}] transitions at 6548~\AA\ and 6584~\AA. This interpretation appears less likely for SN2018cqj/ATLAS18qtd because in both cases it would involve larger blueshifts/redshifts  with respect to the laboratory wavelengths and also there is no theoretical prediction or expectation of detecting [\ion{N}{2}] in SNe~Ia. 

The stripped hydrogen mass estimates using the models of \citet{botyanszki18} for SN2018cqj/ATLAS18qtd are comparable to the estimates obtained for SN2018fhw/ ASASSN-18tb \citep{kollmeier19}, but significantly lower than the $M_{H} \approx 0.1-0.5$~M$_{\odot}$ of stripped hydrogen expected in classical single-degenerate models from sub giant, main sequence or red giant companion stars \citep{marietta00,liu12,boehner17}. Also, the host galaxy environments of SN2018fhw/ASASSN-18tb (in a dwarf elliptical galaxy) and SN2018cqj/ATLAS18qtd (possibly in the outer halo of an S0 galaxy) are associated with old stellar populations, consistent with other low-luminosity, fast-declining Type~Ia SNe \citep{gallagher08,panther19}. These properties are very different from the younger, more massive hydrogen-rich companions of single-degenerate systems that have been proposed to explain Type~Ia-CSM events \citep[e.g.,][]{hamuy03,prieto07,dilday12,silverman13}, which are found in star-forming galaxies and have much higher peak and H$\alpha$ luminosities ($L_{H\alpha} \approx 10^{40}$~erg~s$^{-1}$), and also some more normal Type~Ia SNe with time-variable and/or blueshifted Na~I absorption lines detected in their spectra \citep[e.g.,][]{patat07,sternberg11}. Perhaps this points to a different physical origin for the H$\alpha$ in emission detected in the two fast-declining, low-luminosity Type~Ia in the 100IAS survey. In a future paper we will report the statistics of H$\alpha$ detections in the complete sample of SN~Ia with nebular-phase spectra from 100IAS survey, which will hopefully give us interesting observational clues into the physical origin of the H$\alpha$ detected in these systems.

\acknowledgments

JLP would like to dedicate this article to Leo. We thank M. Tucker for discussions and comments. Support for JLP is provided in part by FONDECYT through the grant 1191038 and by the Ministry of Economy, Development, and Tourism’s Millennium Science Initiative through grant IC120009, awarded to The Millennium Institute of Astrophysics, MAS. PC and SD acknowledge Project 11573003 supported by NSFC. AGY research is supported by the EU via ERC grant No. 725161, the ISF GW excellence center, an IMOS space infrastructure grant and the BSF Transformative program as well as The Benoziyo Endowment Fund for the Advancement of Science, the Deloro Institute for Advanced Research in Space and Optics, The Veronika A. Rabl Physics Discretionary Fund, Paul and Tina Gardner and the WIS-CIT joint research grant; AGY is the recipient of the Helen and Martin Kimmel Award for Innovative Investigation.

Based on observations collected at the European Southern Observatory under ESO programmes 0102.D-0287(A) and 2103.D-5008(A). This research uses data obtained through the Telescope Access Program (TAP), which has been funded by the National Astronomical Observatories of China, the Chinese Academy of Sciences, and the Special Fund for Astronomy from the Ministry of Finance. This research has made use of the NASA/IPAC Extragalactic Database (NED), which is operated by the Jet Propulsion Laboratory, California Institute of Technology, under contract with the National Aeronautics and Space Administration. We acknowledge the usage of the HyperLeda database (http://leda.univ-lyon1.fr). 

ASAS-SN is supported by the Gordon and Betty Moore Foundation through grant GBMF5490 to the Ohio State University and NSF grant AST-1515927. Development of ASAS-SN has been supported by NSF grant AST-0908816, the Mt. Cuba Astronomical Foundation, the Center for Cosmology and AstroParticle Physics at the Ohio State University, the Chinese Academy of Sciences South America Center for Astronomy (CASSACA), the Villum Foundation, and George Skestos.

\software{DoPHOT \citep{dophot1,dophot2}, ISIS \citep{isis1,isis2}, SNooPy \citep{snoopy}, APLpy \citep{robitaille12}, SNID \citep{snid}, IRAF \cite{iraf1,iraf2}}

\begin{deluxetable}{lcccccc}
\tablecolumns{6}
\tablewidth{0pt}
\tablecaption{$BVgri$ photometric measurements of the Type~Ia SN2018cqj/ATLAS18qtd.}
\tablehead{
\colhead{JD} &
\colhead{$B$} &
\colhead{$V$} & 
\colhead{$g$} &
\colhead{$r$} & 
\colhead{$i$} & 
\colhead{Telescope/Instrument} 
}
\startdata
2458291.467 & \ldots  & \ldots  & 16.57(0.09)  & \ldots  & \ldots & ASAS-SN \\
2458293.463 & \ldots  & \ldots  & 16.51(0.07)  & \ldots  & \ldots & ASAS-SN \\
2458295.850 & 16.54(0.04)  & 16.21(0.03)  & 16.38(0.04)  & \ldots  & \ldots & LCOGT~1m \\
2458296.452 & \ldots  & \ldots  & 16.54(0.12)  & \ldots  & \ldots & ASAS-SN \\
2458297.452 & \ldots  & \ldots  & 16.55(0.11)  & \ldots  & \ldots & ASAS-SN \\
2458297.492 & 16.64(0.05)  & 16.23(0.04)  & 16.44(0.05)  & 16.11(0.03)  & 16.74(0.04) & LCOGT~1m \\
2458298.449 & \ldots  & \ldots  & 16.74(0.12)  & \ldots  & \ldots & ASAS-SN \\
2458299.855 & 16.93(0.04)  & 16.27(0.02)  & 16.57(0.03)  & 16.21(0.02)  & 16.70(0.03) & LCOGT~1m\\
2458303.465 & 17.51(0.06)  & 16.50(0.04)  & 17.00(0.04)  & 16.42(0.04)  & 16.95(0.05) & LCOGT~1m\\
2458309.457 & 18.38(0.10)  & 16.99(0.05)  & 17.90(0.09)  & 16.71(0.04)  & 17.04(0.08) & LCOGT~1m \\
2458490.881 & \ldots  &  22.14(0.05) & \ldots & \ldots  & \ldots & VLT~UT1/FORS2\\
2458546.618 & \ldots  &  22.90(0.15) & \ldots & \ldots  & \ldots & du~Pont/WFCCD\\
2458582.586 & \ldots  &  23.28(0.15) & \ldots & \ldots  & \ldots & du~Pont/WFCCD\\
2458607.018 & \ldots  &  23.95(0.10) & \ldots & \ldots  & \ldots & VLT~UT1/FORS2 \\
\enddata
\tablecomments{The values in parenthesis are the $1\sigma$ error bars in the measurements. The magnitudes in the $BV$ filters are given in the Vega system and the magnitudes in the $gri$ filters are in the AB system.}
\label{table:tab1}
\end{deluxetable}

\begin{deluxetable}{lccccc}
\tablecolumns{6}
\tablewidth{0pt}
\tablecaption{Summary of spectroscopic observations of the Type~Ia SN2018cqj/ATLAS18qtd.}
\tablehead{
\colhead{JD} &
\colhead{Wavelength range (\AA)} &
\colhead{Resolution (\AA)} & 
\colhead{Exposure time (sec)} &
\colhead{Airmass} & 
\colhead{Telescope/Instrument} 
}
\startdata
2458314.5 & $3800-9300$   & 8.2  & 160   &  2.3  & du~Pont/WFCCD\\
2458490.9 & $4200-9600$   & 9.9  & 7040  &  $1.1-1.4$ & VLT~UT1/FORS2 \\
2458607.5 & $4200-9600$   & 9.9  & 5800  &  $1.0-1.4$ & VLT~UT1/FORS2 \\
\enddata
\tablecomments{The spectral resolution is measured as the FWHM of the [\ion{O}{1}]~5577~\AA\ sky line.}
\label{table:tab2}
\end{deluxetable}


\end{document}